\documentclass[a4paper,11pt]{article}
\usepackage{jcappub} 
\usepackage{lineno}

\usepackage{siunitx}
\usepackage{physics}
\usepackage{subcaption}
\usepackage{placeins}

\abstract{

The reciprocity approach is a powerful method to determine the expected signal power of axion haloscopes in a model-independent way. Especially for open and broadband setups like the MADMAX dielectric haloscope the sensitivity to the axion field is difficult to calibrate since they do not allow discrete eigenmode analysis and are optically too large to fully simulate. The central idea of the reciprocity approach is to measure a reflection-induced test field in the setup instead of trying to simulate the axion-induced field. In this article, the reciprocity approach is used to determine the expected signal power of a dish antenna and a minimal dielectric haloscope directly from measurements. The results match expectations from simulation but also include important systematic effects that are too difficult to simulate. In particular, the effect of antenna standing waves and higher order mode perturbations can be quantified for the first time in a dielectric haloscope. 
}

\begin{document}

\title{Experimental determination of axion signal power of dish antennas and dielectric haloscopes using the reciprocity approach}

\author[a]{J. Egge,\footnote{Corresponding author.}}

\author[a]{M. Ekmedžić,}

\author[a]{A. Gardikiotis,\footnote{Now at INFN, Sezione di Padova, Padova, Italy.}}
\author[a]{E. Garutti,}

\author[d]{S. Heyminck,}
\author[d]{C. Kasemann,}
\author[e]{S. Knirck,}
\author[d]{M. Kramer,}

\author[a]{C. Krieger,}
\author[b]{D. Leppla-Weber,}

\author[a]{S. Martens,}
\author[c]{E. Öz,}

\author[a]{N. Salama,}

\author[c]{A. Schmidt,}
\author[c]{H. Wang,}

\author[d]{G. Wieching}

\affiliation[a]{Universität Hamburg,
22761 Hamburg,
Germany
}
\affiliation[b]{Deutsches Elektronen-Synchrotron DESY, Germany}

\affiliation[c]{RWTH Aachen,
52074 Aachen,
Germany
}
\affiliation[d]{Max Planck Institute for Radio Astronomy, 
53121 Bonn, 
Germany}

\affiliation[e]{Fermi National Accelerator Laboratory, Batavia, IL 60510, USA}

\emailAdd{jacob.egge@uni-hamburg.de}


\date{\today}


\maketitle
\flushbottom

\section{\label{sec:intro}Introduction}
The axion is an increasingly popular particle candidate for dark matter with more and more experiments trying to detect it. Axions are a consequence of the hypothetical Peccei-Quinn mechanism that explains the absence of charge-parity (CP) violations in the strong interaction, also known as the strong CP problem \cite{peccei_quinn1977,wilczek1978,weinberg1978}. If axions exist, they would not only solve the strong CP problem but could also be produced in the early universe in sufficient abundance to constitute all of dark matter \cite{presikill1983,abott1983,dine1983}. Its mass $m_a$ and coupling to standard model particles is, however, largely unconstrained and experiments have to cover many orders of magnitude\cite{semertzidis2022}. Experiments that search for axions from the galactic dark matter halo are called haloscopes and most of them do so via the coupling of axions to photons which is effectively described by axion-modified Maxwell's equations \cite{sikivie1983}. Together with a strong and static external magnetic (electric) field $\vb*{B}_e$ ($\vb*{E}_e$), the axion field $a$ sources an effective current density $\vb*{J}_a$ that enters Ampère's law

\begin{equation}
\vb*{J}_a = \frac{g_{a\gamma}}{Z_0} \qty(\dot{a} \vb*{B}_e -\vb*{E}_e \cross \grad{a}), 
\label{Eq:axion_current}
\end{equation}

where $g_{a\gamma}$ is the axion-photon coupling constant and $Z_0$ the impedance of free space. Throughout this article, SI units are used. Assuming negligible axion momentum, $\vb*{J}_a$ oscillates with frequency $\nu_a \sim m_a$ and in turn sources the axion-induced fields $(\vb*{E}_a, \vb*{H}_a)$. The traditional haloscope design is a cavity with resonance frequency $\nu_a$ such that the axion-induced fields are amplified and detected as a power excess above background \cite{admx_analysis2021,capp2021}. In the effort to reach unexplored parameter space, new approaches are being developed that deviate from the traditional cavity haloscope. This is because the cavity volume $V$ and resonance frequency $\nu_a$ are tightly coupled by $V \sim \nu_a^{-3}$ making the cavity impractically large for small axion masses $m_a$ or reducing the signal power, proportional to $V$, too much for large $m_a$. Thus, a lot of new approaches can be thought of as attempting to decouple $V$ from $m_a$. One way to do this is open systems like dish antennas \cite{horns2013,brass2023,dosuerr2023} or dielectric haloscopes \cite{caldwell2017} where the open boundary conditions, to first order, allow scaling $V$ independently of $\nu_a$. A drawback is that these optically large setups become very hard to simulate. Simulations are, however, required to predict the axion-induced fields $(\vb*{E}_a, \vb*{H}_a)$ which cannot be measured in the absence of an axion discovery. To not entirely rely on simulations, one can study the response of the setup to a different source current density $\vb*{J}_R$, for example from a signal generator in a reflection measurement. This then excites the reflection-induced fields $(\vb*{E}_R, \vb*{H}_R)$. Closed resonators usually allow for only a single, discrete mode at a given frequency and, up to normalization, $(\vb*{E}_a, \vb*{H}_a)$ and $(\vb*{E}_R, \vb*{H}_R)$ will be the same. In contrast, open systems allow infinitely many continuous modes at the same frequency and the axion-induced and reflection-induced fields are not the same. Here, the reciprocity approach enters where the axion signal power $P_{\mathrm{sig}}$, instead of depending on the axion-induced fields, is rewritten to depend on the reflection-induced fields that are excited with input power $P_{\mathrm{in}}$ \cite{Egge_2023}, 
\begin{equation}
    P_{\mathrm{sig}} = \frac{g_{a\gamma}^2}{16 Z_0^2P_{\mathrm{in}}}\abs{\int_{V_a} \dd{V} \vb*{E}_R \vdot  \left(\dot{a} \vb*{B}_e -\vb*{E}_e \cross \grad{a} \right)}^2.
    \label{Eq:power_ax_reciprocity}
\end{equation} 
The axion signal power $P_{\mathrm{sig}}$ now only depends on measurable quantities directly controlled by the experiment and the axion parameters one wishes to constrain. The external fields are confined to the conversion volume $V_a$ in the sense that the contribution to $P_{\mathrm{sig}}$ from outside $V_a$ becomes negligible. Reciprocity is a property of linear systems so in addition to the usual requirements of linear and time-independent material properties an important assumption is that the axion-modified Maxwell's equations are linearized for strong external fields and small axion-photon coupling \cite{rodd2021}. Another assumption is that the reflection-induced fields are excited via the same port that is used to measure the axion-induced fields. This port must support only a single mode whereas the rest of the optical system can be completely arbitrary.  
 
While the reciprocity approach was validated numerically before \cite{schulz2018, Egge_2023}, it has not been used experimentally. In this article, this will be achieved on a MADMAX dielectric haloscope test setup, measuring the expected axion signal power of an open haloscope for the first time. A key ingredient for this is non-resonant perturbation theory, more commonly known as the non-resonant bead pull method, which is used to measure the reflection-induced fields \cite{steele1966}.
The article is structured as follows: In section \ref{sec:setup} the working principles of a dielectric haloscope in general and the experimental setup specifically are described. Section \ref{sec:beadpull} introduces non-resonant perturbation theory and its application on measuring $\vb*{E}_R$. The simulations used in this study are briefly described in section \ref{sec:simulation}. The results are presented in section \ref{sec:results}. This is done by successively building up the dielectric haloscope, validating the measurement procedure on the simplest setup first before finally measuring $P_{\mathrm{sig}}$ on a minimal dielectric haloscope. Conclusions are then drawn in section \ref{sec:conclusion}.

\section{Minimal dielectric haloscope}
\label{sec:setup}
\begin{figure}
    \centering
    \begin{subfigure}[b]{0.49\textwidth}
        \centering
        \includegraphics[width=\textwidth]{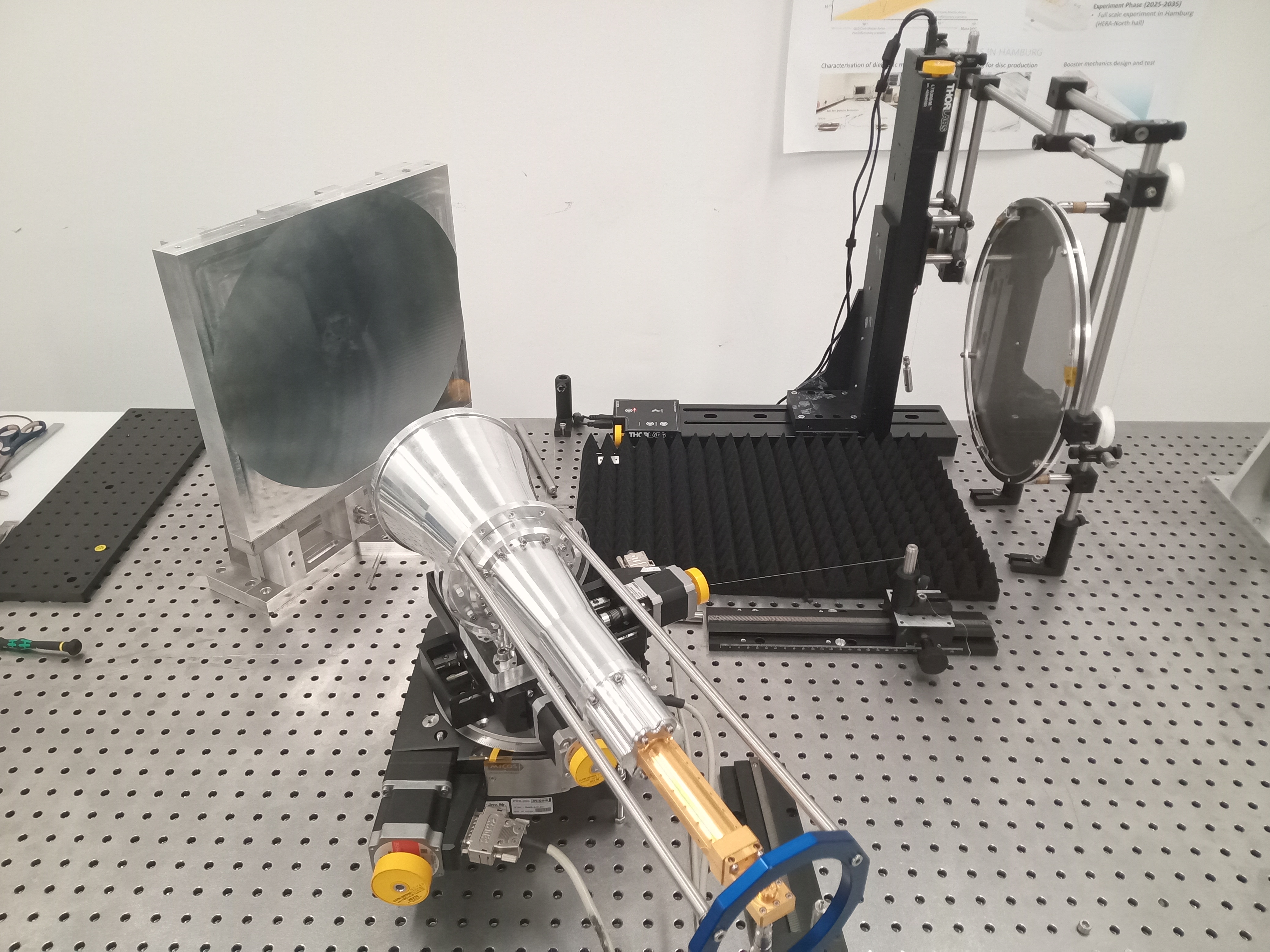}
        \label{fig:setup_pic}
    \end{subfigure}
    \hfill
    \begin{subfigure}[b]{0.49\textwidth}
        \centering
        \includegraphics[width=\textwidth]{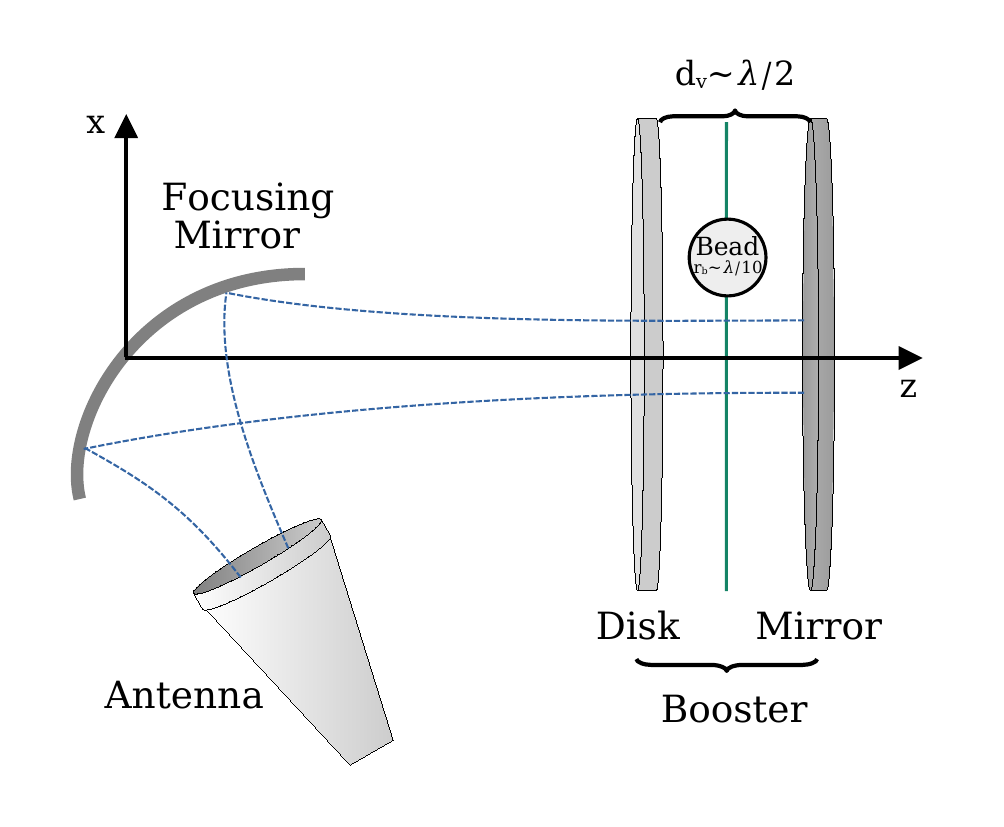}
        \label{fig:setup_sketch}
    \end{subfigure}
    \caption{Minimal dielectric haloscope setup consisting of an antenna, focusing mirror, and booster. The antenna is connected to a VNA that measures the reflection coefficient $\Gamma$. A dielectric bead can be moved in between or in front of the booster in all three dimensions to map the electric field.}
    \label{fig:setup}
\end{figure}

A dielectric haloscope like MADMAX uses many dielectric disks to boost the axion signal to detectable levels \cite{caldwell2017}. From the perspective of the axion-induced fields, each interface emits a traveling wave along its normal direction. These emissions can then interfere constructively with each other as well as resonate between the disks, boosting the total signal power. A plane mirror at one end of the dielectric stack ensures that all emissions travel toward the receiver system. The stack of dielectric disks plus a plane mirror is called a booster. The power boost factor $\beta^2$ is the amplification compared to the power $P_0$ emitted from just a perfect plane mirror with area $A$,
\begin{gather}
    \beta^2 \equiv \frac{P_{\mathrm{sig}}}{P_0} \\
    P_0 = \frac{g_{a\gamma}^2 c^2 \abs{a_0}^2 \abs{\vb*{B}_e}^2 A}{2Z_0}.
    \label{Eq:boost_factor}
\end{gather}
The booster of a minimal dielectric haloscope, shown in figure \ref{fig:setup}, consists of one dielectric disk and a plane mirror. A focusing mirror couples the axion-induced fields to an antenna connected to the receiver chain that then measures $P_{\mathrm{sig}}$.

In the reciprocity approach, the vector network analyzer\footnote{Rhode \& Schwarz ZVA67} (VNA) represents $\vb*{J}_R$ and sources the reflection-induced field $\vb*{E}_R$. From here on out, the subscript differentiating between reflection-induced and axion-induced fields is dropped and $\vb*{E}$ refers to the reflection-induced fields unless stated otherwise. The antenna emits a beam that is focused onto the booster by the focusing mirror. When the beam enters the booster, it can resonate between the semi-transparent dielectric disk and plane mirror when their separation is roughly half a wavelength. This increases $\vb*{E}$ relative to $P_{\mathrm{in}}$ which then increases $P_{\mathrm{sig}}$ according to equation \eqref{Eq:power_ax_reciprocity}. Inside the disk, $\vb*{E}$ has a negative lobe which is suppressed by the disk's dielectric constant $\epsilon$ which further increases $P_{\mathrm{sig}}$. This describes the basic working principle of a dielectric haloscope in the longitudinal direction. For a complete description and ultimately full signal power, the behavior in the transverse direction is important. 

The coordinate system $(x,y,z)$ is defined as follows: The origin is the reflection point of the focusing mirror. The optical axis points to the center of the booster and defines the z-axis. The y-axis points up from the optical table and the x-axis points towards the wall in figure \ref{fig:setup}. The external magnetic field $\vb*{B}_e$ is imagined to point along the y-direction. It is assumed to be homogeneous over the extent of the booster and decreasing to negligible magnitude outside the booster. The focusing mirror and antenna would not be magnetized. The circular horn antenna produces a near-perfect Gaussian beam, polarized in y-direction, that is redirected and focused by the off-axis ellipsoidal focusing mirror. The result, ideally, is again a Gaussian beam converging onto the booster. Gaussian beams are the fundamental mode of the Hermite-Gaussian modes 

\begin{equation}
 \begin{aligned}
E_{ml}(x,y,z') &=E_{0} \frac{w_{0}}{w(z)} H_{l} \left(\frac{\sqrt{2} x}{w(z)}\right) H_{m} \left(\frac{\sqrt{2} y}{w(z)}\right)  \exp \left({-{\frac {r^2}{w(z)^{2}}}}\right)\\&\times \exp \left(-i\left(kz'+k{\frac {r^{2}}{2R(z')}}-\psi_{ml}(z')\right)\right). 
\end{aligned}   
\label{Eq:hermite_gaussian_modes}
\end{equation}

where $r = \sqrt{x^2 + y^2}$ is the radial distance from the optical axis, $z'=z-z_0$ the axial distance from the beam waist at $z_0$ and of radius $w_0$, $k$ the in-media wave number, and $E_0$ the complex amplitude at $z' = 0$. The first line in equation \eqref{Eq:hermite_gaussian_modes} describes the transverse beam shape while the second line describes the phase evolution. The Hermite polynomials $H_{m,l}$ equal unity for $m,l=0$, recovering the Gaussian beam. The waist radius depends on $z'$ as
\begin{equation}
    w(z') = w_0 \sqrt{1+\left(\frac{z'}{z_R}\right)^2},
\end{equation}
with Rayleigh range $z_R = \frac{k w_0^2}{2}$. The radius of curvature of the beam's wavefronts $R(z')$ is defined as
\begin{equation}
    R(z') = z' \left[1 + \left(\frac{z_R}{z'}\right)^2\right].
\end{equation}
The Gouy phase $\psi_{ml}(z')$ is 
\begin{equation}
    \psi_{ml}(z') = (m+l+1) \arctan{\left(\frac{z'}{z_R}\right)}.
\end{equation}
The waist radius on the booster side is $\sim \SI{45}{\milli \metre}$ at $\SI{20}{\giga\hertz}$ resulting in a Rayleigh range of $z_R \sim \SI{45}{\centi\metre}$. For $z'<z_R$, $\psi_{00}(z')$ is small, $R(z')$ is large, and the phase term of $E_{ml}(x,y,z')$ is well approximated by a plane wave. 

\begin{figure}
    \centering
    \begin{subfigure}[b]{0.49\textwidth}
        \centering
        \includegraphics[width=\textwidth]{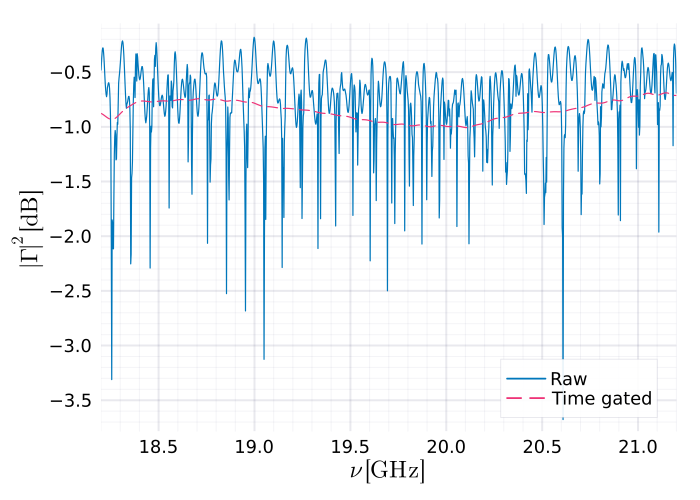}
        \caption{}
        \label{fig:intro_s11_freq}
    \end{subfigure}
    \hfill
    \begin{subfigure}[b]{0.49\textwidth}
        \centering
        \includegraphics[width=\textwidth]{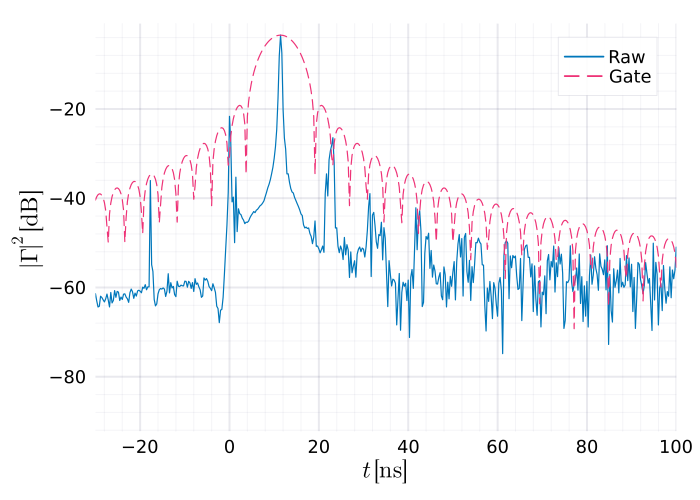}
        \caption{}
        \label{fig:intro_s11_time}
    \end{subfigure}
    \caption{Reflection coefficient $\Gamma$ of the minimal dielectric haloscope. \textbf{(a)}: Magnitude of $\Gamma$ over frequency. The spectrum is dominated by the antenna-booster resonances which are removed via time gating (dashed line). \textbf{(b)}: Fourier transform of (a) a gate (dashed) is applied to filter out antenna-booster resonances.}
    \label{fig:1_disk_s11}
\end{figure}

The booster reflects the antenna beam from more than one interface which even under ideal circumstances are not matched to its curvature $R(z)$. Consequently, the reflected beam is different from the emitted antenna beam. In reality, geometrical inaccuracies like tilts and surface deformations distort the beam further. The distortions can be described as exciting higher order modes (HoM) $E_{m>0,l>0}$. The reflected beam thus contains a fraction of HoMs. Since the HoMs are orthogonal to the near-perfect fundamental mode the antenna emits, they are reflected from the antenna instead of being received. For the HoM, the antenna acts like a mirror and the antenna-booster system is essentially a leaky cavity. The distance between the antenna and the booster is $\mathcal{O}(100 \lambda)$ and so the harmonics of the antenna-booster resonance will be very close in frequency. The effect is seen in the reflection coefficient $\Gamma$ that is measured by the VNA. Its absolute value is plotted in figure \ref{fig:intro_s11_freq}. The spectrum is dominated by the harmonics of the antenna-booster resonance while the actual booster resonance is hidden underneath. 

A common method to filter out undesired reflections from a measurement is time gating. By Fourier transforming $\Gamma$ into the time domain, figure \ref{fig:intro_s11_time}, the antenna-booster resonance manifests itself as successive peaks separated by an interval corresponding to twice the distance between antenna and booster. Applying a gate on the main peak and transforming back into the frequency domain isolates the signal that is not reflected by the antenna, removing the antenna-booster resonances. Time gating relies, however, on the phase information of the signal. Since the axion signal is measured only in power, it cannot be time-gated and antenna reflections cannot simply be ignored. The reciprocity approach offers a model-independent way to measure the expected signal power, regardless of undesired reflections or not. To do that, one needs to measure the reflection-induced electric field $\vb*{E}$ throughout the booster.

\section{Bead pull method}
\label{sec:beadpull}
Non-resonant perturbation theory \cite{steele1966}, also known as bead pull method, is a natural choice for measuring the reflection-induced electric field $\vb*{E}$. It relates the change in reflection coefficient $\Gamma$ from introducing an object, in this case a dielectric bead, into an arbitrary optical system to the electric field at that object's position. In its most general form, one finds

\begin{equation}
\Delta \Gamma = -\frac{i \epsilon_0 \omega (\epsilon_B-1)}{4 P_{\mathrm{in}}} \int_{V_B} \dd{V}\vb*{E}_{B} \cdot \vb*{E},
\label{Eq:steele_exact}
\end{equation}
where $\vb*{E}_{B}$ is the electric field in the presence of the bead with relative permittivity $\epsilon_B$ and the integration is performed over the volume of the bead $V_B$. A background medium of $\epsilon = 1$ is already assumed here.  
While \eqref{Eq:steele_exact} is an exact relation applicable to an arbitrary object of any size, it is not very usable since it includes the electric field inside the bead $\vb*{E}_{B}$. To relate the internal field $\vb*{E}_{B}$ to the exciting field $\vb*{E}$ is a common problem in scattering theory. For a homogeneous dielectric sphere with a radius $r_B$ this can be solved analytically by Mie theory \cite{Bohren1998}. For any radius but $r_B \ll \lambda$ the expressions quickly become unwieldy. In particular, the fact that $\vb*{E}_{B}$ is not homogeneous over the extent of the bead complicates matters significantly as the integrand of equation \eqref{Eq:steele_exact} cannot be separated. A practical approach is then to define the bead factor

\begin{equation}
    \delta_e = \frac{V_B(\epsilon_B-1)}{\alpha_0} \frac{\langle \vb*{E}_{B} \cdot \vb*{E} \rangle}{\vb*{E}^2},
    \label{Eq:beadfactor}
\end{equation}

where the electric field in the denominator is evaluated at the bead's center, the angled brackets denote the average over $V_B$, and $\alpha_0$ is the polarizability of a dielectric sphere in a static, homogeneous electric field

\begin{equation}
    \alpha_0 = 4\pi r_B^3 \frac{\epsilon_B-1}{\epsilon_B+2}.
\end{equation}

Another quantity similar to $\delta_e$ is 
\begin{equation}
    \delta_c = \frac{V_B(\epsilon_B-1)}{\alpha_0} \frac{\langle \vb*{E}_{B} \cdot \vb*{E}^{*} \rangle}{\abs{\vb*{E}}^2},
    \label{Eq:beadfactor_c}
\end{equation}
which will be needed in the dish antenna and dielectric haloscope setup to better deconvolve $\delta_e$ from the measured $\Delta \Gamma$. Generally, $\delta_e$ and $\delta_c$ are functions of both position and frequency. 
In the limit of an infinitely small bead, $\delta_e$ and $\delta_c$ become unity
\begin{equation}
    \lim_{k r_B \to 0} \delta_{e,c} = 1.
\end{equation}
Mie theory is used to compute $\delta_{e,c}$ for finite radii. The analytical expressions are cumbersome and can be found in the appendix \ref{sec:app_mie}. Equation \eqref{Eq:steele_exact} can then be rewritten as
\begin{equation}
\Delta \Gamma = \frac{\epsilon_0 \delta_e \alpha_0 \omega}{4P_{\mathrm{in}}}\vb*{E}^2. 
\label{Eq:nonresonant_small_steele}
\end{equation}

It is worth noting that $\vb*{E}^2$, just as $\Delta \Gamma$, is a complex quantity with phase information. It also contains contributions from all spatial components whereas only the component parallel to $\vb*{B}_e$ will contribute to $P_{\mathrm{sig}}$. Thus an important assumption is that $\vb*{E}$ is polarized along the would-be external magnetic field $\vb*{B}_e$. Since the cross-polarization of the antenna is $<\SI{30}{\deci\bel}$, and the alignment of optical components is well controlled, the contribution of cross-polarization to $\Delta \Gamma$ is expected to be negligible. Finally, and quite conveniently, $\Delta \Gamma$ inversely depends on the input power $P_{\mathrm{in}}$ so that it cancels when inserting equation \eqref{Eq:nonresonant_small_steele} into \eqref{Eq:power_ax_reciprocity}. 

The bead used in this study is made of alumina. Its relative permittivity was measured to be $\epsilon_B = \num{9.23(5)}$ by inserting it into a resonant cavity with known dimensions and observing shifts in resonance frequencies. This suggests some impurities as literature puts alumina closer to $\epsilon \sim 10$. Its radius of $r_B=\SI{1.46(1)}{\milli \metre}$ is much smaller than the beam waist $w_0$ and roughly $\lambda/10$. Consequently, $\vb*{E}$, but not $\vb*{E}_B$, is essentially homogeneous in transverse direction over the extent of the bead while in longitudinal direction small but non-negligible variation is expected. The bead is attached to a string made from high-modulus polyethylene (HMPE) with a diameter of $\SI{0.1}{\milli\metre}$. Judging from the relatively low $\epsilon$ of HMPEs and the tiny volume of the string, a negligible effect of the string on $\Delta \Gamma$ is expected. The bead can be positioned in three dimensions. In x-direction, the string is pulled against a counterweight by a stepper motor with an estimated accuracy of $<\SI{100}{\micro\metre}$. Two linear translation stages\footnote{Thorlabs LTS300} move the whole bead pull setup in y and z-direction with an accuracy of $<\SI{5}{\micro\metre}$.

\section{Simulation}
\label{sec:simulation}
The setups under consideration are simulated using COMSOL Multiphysics\textsuperscript{\tiny\textregistered}, a finite element simulation software. As the setups are optically large, 2D axisymmetry is used to make simulation possible in the first place. This means that only azimuthally symmetric geometry can be simulated. The focusing mirror breaks this symmetry and thus the simulation does not include the focusing mirror and antenna. Instead, the setups are simulated using the scattered field formulation where the background field is set to the Gaussian beam that is expected to be produced by the antenna and focusing mirror. While saving considerable computational cost, the simulation cannot include the antenna-booster resonances. Nevertheless, one can still compare measurement and simulation by filtering out the antenna-booster resonances from the measurement using time gating. 
To simulate the linearly polarized fields produced by the antenna in 2D axisymmetric mode, all relevant field components $X$ are decomposed into a left and right circularly polarized part \cite{Knirck_2019}:
\begin{equation}
    X(r,\phi,z) =  X^+(r,z)e^{-i \phi} +  X^-(r,z)e^{+i \phi}
    \label{Eq:azimuthal_decomp}
\end{equation}
Only the azimuthally symmetric components $X^{\pm}(r,z)$ are simulated and later added according to \eqref{Eq:azimuthal_decomp}. 
The entire domain is surrounded by perfectly matched layers to absorb any outgoing radiation. The dish antenna setup is simulated by adding a plane mirror using an impedance boundary condition. The axion signal power is calculated using \eqref{Eq:power_ax_reciprocity}. For the dielectric haloscope, a disk is added with the appropriate dimensions. 


\section{\label{sec:results}Results}
\FloatBarrier

The reciprocity approach is tested on three different setups. The first setup consists of just the antenna and focusing mirror and characterizes the antenna beam as well as bead pull systematics. Next, a plane mirror is added. This configuration is also known as a dish antenna. Finally, a dielectric disk completes the minimal dielectric haloscope. The input power $P_{\mathrm{in}}$ is arbitrarily set to $\SI{1}{\watt}$ as its true value,  which is not known precisely, cancels in the end. As $\vb*{E}$ and $\vb*{B_e}$ are assumed to be polarized in the y-direction, it is sufficient to use the scalar fields $E = \vb*{E} \vdot \vb*{y}$ and $B_e = \vb*{B}_e \vdot \vb*{y}$. 

\subsection{Antenna beam}
\begin{figure}
    \centering
    \includegraphics[width=\textwidth]{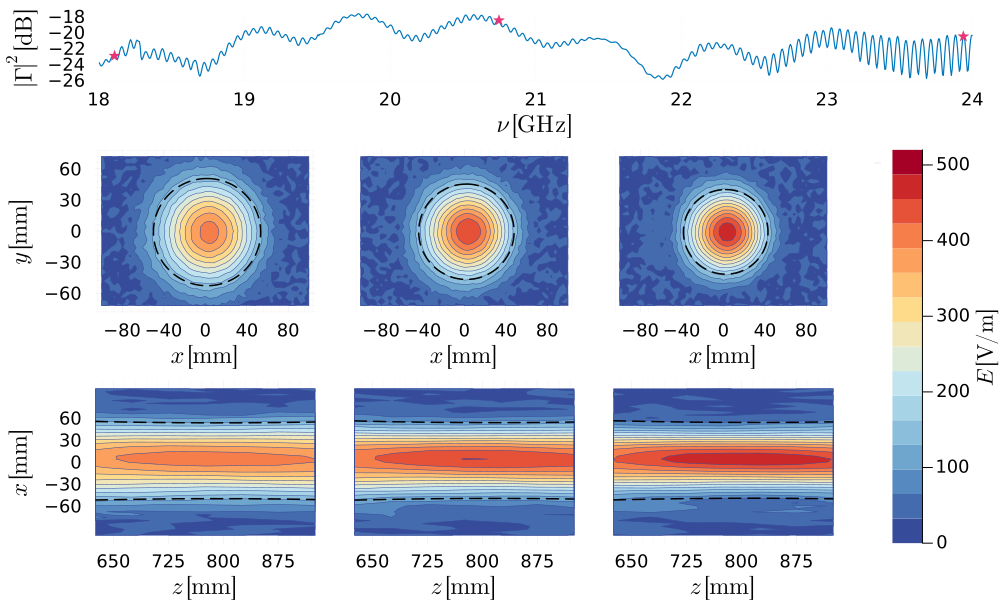}
    \caption{Electric field $E$ produced by the antenna and focusing mirror. \textbf{Top}: $|\Gamma|^2$ over frequency. The markers indicate the frequencies at which $E$ is shown below from left to right. \textbf{Middle row}: transverse beam profile at three different frequencies. \textbf{Bottom row}: Longitudinal beam profile. The dashed lines indicate the beam waist position and size.}
    \label{fig:no_booster_E_evol}
\end{figure}
Before measuring $P_{\mathrm{sig}}$ on the minimal dielectric haloscope, the bead pull method is tested on a simpler setup consisting of just the antenna and focusing mirror. The relevant antenna and focusing mirror characteristics are known from independent measurements and ${E}$ can be benchmarked against expectations. Additionally, the measured beam parameters will serve as an input to simulations. RF absorbers are put in place of the booster to emulate free space. The reflection coefficient $\Gamma$ for this setup can be seen in the top panel of figure \ref{fig:no_booster_E_evol}. The non-zero baseline and long oscillations come from reflections at the antenna input. The short oscillations come from reflections at the RF absorbers. These reflections are irrelevant as the absorbers will later be replaced by the booster. Thus $\Gamma$ is time-gated to exclude absorber reflections. The bead is then moved across the beam at different heights and distances from the focusing mirror. Measurements of $\Gamma$ where the bead is well outside the beam are used as a reference for $\Delta \Gamma$ from which ${E}$ is calculated using \eqref{Eq:nonresonant_small_steele}. The detailed procedure that accounts for signal drift, as well as potential branching issues when taking the complex square root of $\Delta \Gamma$, is described in the appendix \ref{sec:appendix}. A bead factor of $\delta_e=1$ is assumed for now. Deviations from this assumption are discussed later. 

Figure \ref{fig:no_booster_E_evol} shows the measured beam in the $xy$ and $xz$ plane at three different frequencies as indicated by the markers in the spectrum above. The dashed line shows the beam waist $w(z)$ that was extracted by fitting a Gaussian beam ${E}_{00}$ separately for each frequency to the measured field. The overlap with the fundamental mode is found to be $>\SI{96}{\percent}$.

\begin{figure}
    \centering    
    \begin{subfigure}[b]{0.49\textwidth}
        \centering
        \includegraphics[width=\textwidth]{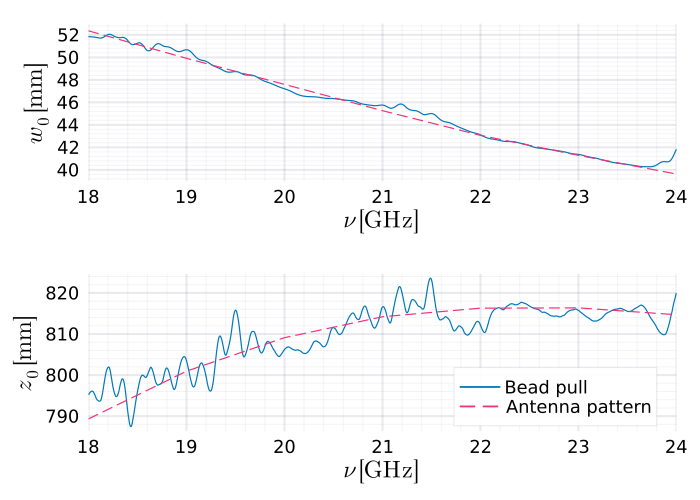}
        \caption{}
        \label{fig:no_booster_w0_z0}
    \end{subfigure}   
    \hfill   
    \begin{subfigure}[b]{0.49\textwidth}
        \centering
        \includegraphics[width=\textwidth]{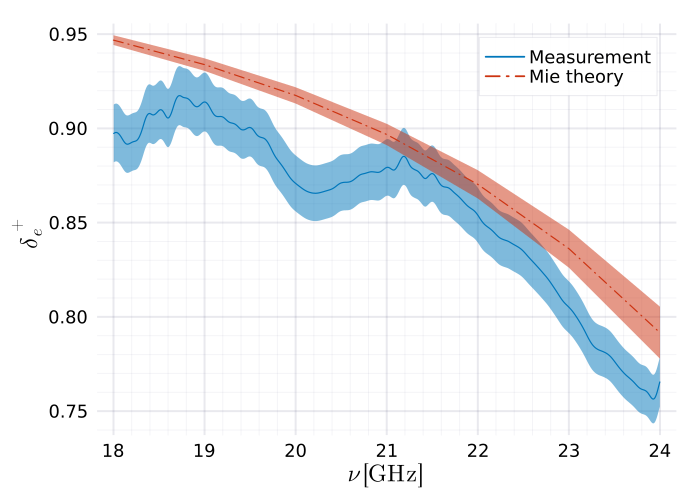}
        \caption{}
        \label{fig:no_booster_beadfactor}
    \end{subfigure}
    \caption{\textbf{(a)}: Beam waist size $w_0$ (top panel) and position $z_0$ (bottom panel) over frequency as measured by the bead pull method (solid line). The dashed line shows the expectation from the antenna and focusing mirror characteristics. \textbf{(b)}: Bead factor $\delta_e$ from measurement (solid line) and FEM simulation (dashed line).}
    \label{fig:no_booster_waist}
\end{figure}

The fitted beam waist size $w_0$ and position $z_0$ are shown in figure \ref{fig:no_booster_w0_z0} together with the expectation from the antenna radiation pattern that was measured independently in a reverberation chamber by the Max Planck Institute for Radio Astronomy who also designed the antenna. The radiation pattern gives the frequency-dependent beam waist size $w_A$ and position $z_A$ on the antenna side which together with the focal length $f$ of the focusing mirror can be used to calculate $w_0$ and $z_0$ via the thin lens equation,

\begin{equation}
    \frac{w_A^2}{w_0^2} = \frac{z_A - f}{z_0 - f}.
    \label{Eq:thin_lens}
\end{equation}
The antenna radiation pattern gives $z_A$ relative to the antenna's aperture whereas for equation \eqref{Eq:thin_lens} the absolute distance to the reflection point of the focusing mirror is needed which is not known precisely. To obtain the good match of figure \ref{fig:no_booster_w0_z0}, a frequency-independent offset of $\sim \SI{6}{\milli \meter}$ is added to $z_A$, still within the estimated uncertainty on the absolute positions of antenna, focusing mirror and bead. The focal length also needs to be scaled by $\SI{4.8}{\percent}$ from the geometric value of $f = \SI{352.5}{\milli \meter}$. This is a bit surprising since the focusing mirror was machined with $\sim \SI{10}{\micro\metre}$ precision. Alternatively, one could scale both $z_A$ and $z_0$ which in equation \eqref{Eq:thin_lens} is equivalent to changing the focal length. However, the change in distances is on the order of $\si{\centi\metre}$ which seems excessive. As $w_0$ and $z_0$ are extracted from a fit, they somewhat depend on the exact fitting procedure. Fine-tuning the procedure can only improve the mismatch in focal length by roughly $\SI{1}{\percent}$. The measured Gaussian beam might also be systematically broadened by the bead pull measurements. The finite size of the bead would do this but its effect is orders of magnitude smaller than the observed mismatch since $r_B \ll w_0$. Similarly, a lack of accuracy in motor step size seems not large enough. Overall, it remains most likely that a combination of inaccuracies of absolute distances, misalignment, and focusing mirror quality leads to a different beam than expected. Nonetheless, it is worth considering the worst-case scenario where a systematic broadening would lead to an overestimation of $P_{\mathrm{sig}}$ by improving the overlap to the uniform axion field. For a Gaussian beam, $P_{\mathrm{sig}}$ scales with $w_0^2$. The mismatch between $w_0$ from the bead-pull measurements and the naive expectation from the antenna pattern then leads to a potential overestimation of $P_{\mathrm{sig}}$ of $\sim \SI{15}{\percent}$. Allowing a realistic deviation in absolute distances and focal length of $\sim \SI{1}{\percent}$, the potential overestimation of $P_{\mathrm{sig}}$ reduces to $\sim \SI{7}{\percent}$. 

So far $\delta_e=1$ was assumed. The axion signal power $P_{\mathrm{sig}}$, however, depends on the true value of $E$ and thus $\delta_e$. Over the extend of the bead, ${E}$ can be assumed to be a plane wave for which $\delta_e$ is independent of bead position. Consequently, the shape of $E$ should not be affected by $\delta_e$. In this simple setup without a booster, the power contained in the beam $P_{\mathrm{beam}} \sim P_{\mathrm{in}}$ since inefficiencies coming from the antenna and focusing mirror are negligible. For the antenna, this is apparent from figure \ref{fig:no_booster_E_evol} where $|\Gamma|^2<\SI{-18}{\deci\bel}$. Focusing mirror inefficiencies come from the conduction loss of aluminium and losses due to surface roughness which can be estimated by Ruze's equation \cite{ruze1966}. Using a conductivity of $\sigma_{Al}=\SI{5e7}{\siemens\per\metre}$ and surface errors of $\sigma_{RMS} \sim \SI{10}{\micro \metre}$ yields losses of less than $\SI{-30}{\deci\bel}$ and $\SI{-40}{\deci\bel}$, respectively. $P_{\mathrm{beam}}$ can be calculated by integrating the measured ${E}$ in the $xy$ plane:
\begin{equation}
   P_{\mathrm{beam}} = \frac{1}{2 Z_0} \int \dd{A} \left\lvert {E} \right\rvert^2.  
\end{equation}

Setting $P_{\mathrm{beam}} = P_{\mathrm{in}}$, one then finds from equation \eqref{Eq:nonresonant_small_steele}

\begin{equation}
    \delta_e^{+} = \frac{2}{Z_0 \alpha_0 \omega}\int \dd{A} \left\lvert \Delta \Gamma\right\rvert,
    \label{Eq:beadfactor_measurement}
\end{equation}

where the + in superscript indicates the assumption of a plane wave traveling in the positive z-direction. Figure \ref{fig:no_booster_beadfactor} shows $\delta_e^{+}$ obtained from measurement via \eqref{Eq:beadfactor_measurement} and analytically from Mie theory, see equation \eqref{Eq:beadfactor_mie} in the appendix. The shaded ribbon is the uncertainty on the measured $\delta_e^{+}$ that is in part estimated from observed fluctuations in the measured $\Delta \Gamma$ but largely due to uncertainty in $\alpha_0$ which is in turn dominated by uncertainty in the bead radius $r_B$. The agreement with the result from Mie theory is better than $\sim \SI{8}{\percent}$. The uncertainty from Mie theory again stems largely from the uncertainty in the bead's radius. For now, the remaining mismatch between measurement and simulation will be treated as a systematic uncertainty on $\delta_e$ that later directly translates to an uncertainty on $P_{\mathrm{sig}}$.

\subsection{Dish antenna}
\begin{figure}
    \centering
    \includegraphics[width=\textwidth]{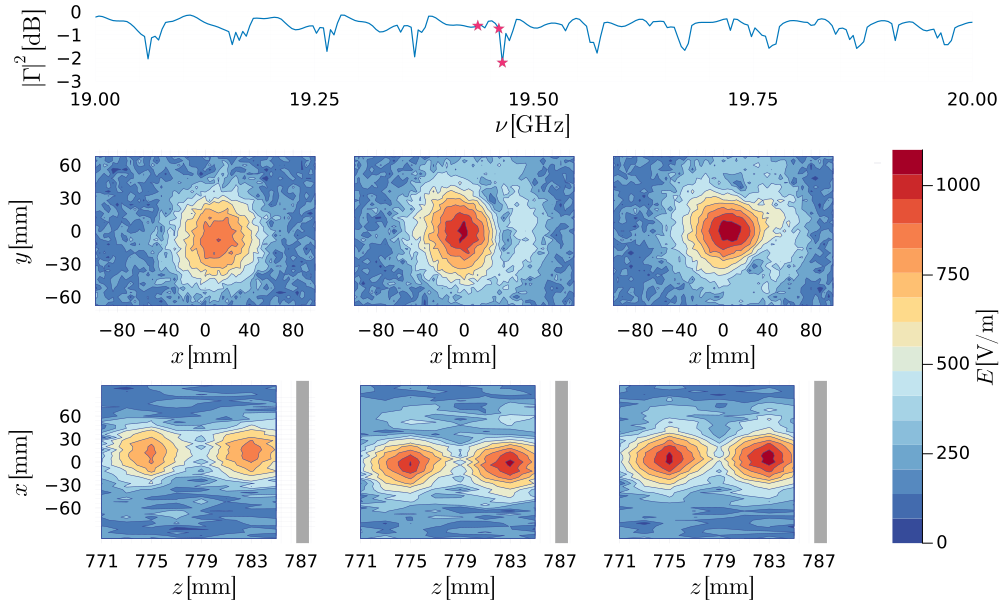}
    \caption{Electric field in front of the dish antenna with a diameter of $\SI{300}{\milli\metre}$ positioned at $z \sim \SI{787}{\milli\metre}$. \textbf{Top panel:} Reflection coefficient of the setup. The markers indicate the frequencies at which the electric field is shown below from left to right. \textbf{Middle row}: Transverse field profile at $z = \SI{783}{\milli\metre}$ with increasing amount of distortion due to antenna-booster resonances. \textbf{Bottom row}: Corresponding longitudinal field profile at $y = \SI{0}{\milli\metre}$ with visible standing wave pattern. The plane mirror is shown as a dark gray band.}
    \label{fig:mirror_E_evol}
\end{figure}

Now that the bead pull method has been benchmarked with a known and relatively simple electric field, the next step is to add a plane mirror forming a dish antenna haloscope. Without any magnetized interface, there is no axion-photon conversion as integrating a traveling wave per equation \eqref{Eq:power_ax_reciprocity} evaluates to zero for a homogeneous ${B}_e$. By adding the plane mirror, ${E}$ becomes a standing wave for which the integral in \eqref{Eq:power_ax_reciprocity} is non-zero. The plane mirror is put at $z_m = \SI{786.5(1)}{\milli\metre}$ perpendicular to the beam propagation and has a diameter of $\SI{300}{\milli\metre}$. Its position is estimated from the linear motor stage position by moving the bead as close as possible to the mirror. As discussed before, distortions of the beam from its original Gaussian shape lead to antenna-booster resonances. Figure \ref{fig:mirror_E_evol} shows the magnitude of ${E}$ in the $xy$ and $xz$ plane as measured in front of the plane mirror at three different frequencies. Off antenna-booster resonance, where $\abs{\Gamma}$ is maximal, the field appears relatively undisturbed from the Gaussian beam while on resonance, low $\abs{\Gamma}$, the contribution from HoMs is clearly visible. Having measured ${E}$, one can now expect two opposing effects that the antenna-booster resonances have on $P_{\mathrm{sig}}$. First, the overall magnitude of ${E}$ increases on resonance which can boost the signal power. Second, the higher order modes generally have a poor overlap with a uniform $a {B}_e$ which would decrease $P_{\mathrm{sig}}$. The z-dependence of ${E}$ can be approximated by plane wave propagation since the measurement range is smaller than the Rayleigh range of $\sim \SI{45}{\centi \meter}$. By integrating ${E}$ in the transverse direction, the analysis can be reduced to one dimension. The one-dimensional electric field amplitude is defined as

\begin{equation}
    E_{1D} = \int \dd{A} {E}.
\label{Eq:intdAE}
\end{equation}
Note that $E_{1D}$ has dimensions of $\si{\volt \metre}$. Crucially, the \textit{complex} electric field $E$ is integrated in \eqref{Eq:intdAE}. With the plane mirror at one end, $E_{1D}$ is a superposition of two plane waves traveling in opposite directions,
\begin{equation}
    E_{1D} = E_{+} e^{-ikz} + E_{-} e^{+ikz}.
    \label{Eq:standing_wave}
\end{equation}

\begin{figure}
    \centering    
    \begin{subfigure}[b]{0.49\textwidth}
        \centering
        \includegraphics[width=\textwidth]{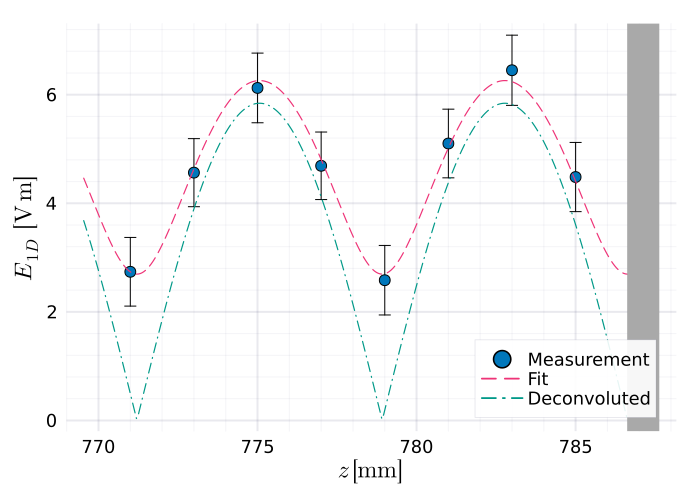}
        \caption{}
        \label{fig:mirror_intdA_E}
    \end{subfigure}    
    \hfill
    \begin{subfigure}[b]{0.49\textwidth}
        \centering
        \includegraphics[width=\textwidth]{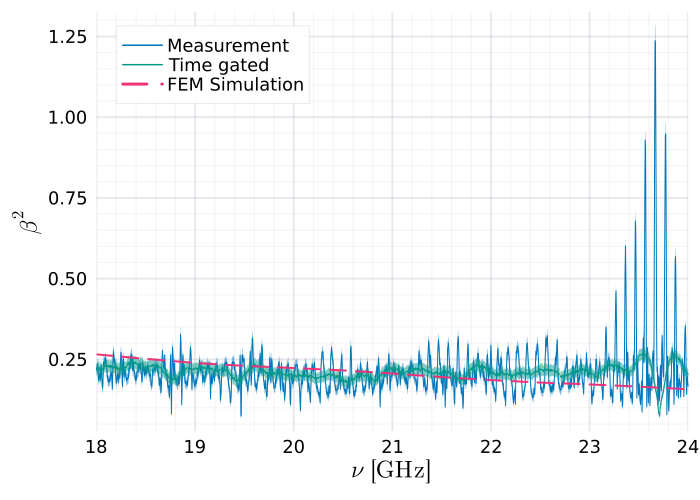}
        \caption{}
        \label{fig:mirror_bf}
    \end{subfigure}
    \caption{\textbf{(a)}: Transversely integrated electric field over $z$ at $\nu \sim \SI{19.44}{\giga\hertz}$. The measurements are fitted by equation \eqref{Eq:fit_fun_delta} (dashed line) allowing to deconvolve finite size effects of the bead from the measurement (dashed-dotted line). \textbf{(b)} Boost factor $\beta^2$ of the dish antenna from measurement and simulation.}
    \label{fig:mirror_integration}
\end{figure}

Figure \ref{fig:mirror_intdA_E} shows the measured $E_{1D}$ over z at a selected frequency. The errors are estimated from fluctuations of $\Delta \Gamma$ at the edges of the booster. The measurement is clearly similar to a standing wave with the exception that at the nodes a non-zero offset remains. The finite size of the bead effectively averages over the extent of the bead, similar to a moving average. This is accounted for by the bead factor $\delta_e$. To obtain the true magnitude of $E_{1D}$, the measured $E_{1D}$ has to be deconvolved from $\delta_e$ which is now a function of both frequency and position. The quantity that is actually measured is $\Delta \Gamma$. Using \eqref{Eq:standing_wave} and the fact that for a single plane wave $\delta_e^{+}$ is independent of position and direction, $\Delta \Gamma$ can be written as

\begin{equation}
    \frac{4 P_{\mathrm{in}} \Delta \Gamma}{i \omega \alpha_0} = \delta_e^{+} \left(E_{+}^2 e^{-2ikz} + E_{-}^2 e^{+2ikz}\right) + 2 \delta_c^{+} E_{+}E_{-},
    \label{Eq:fit_fun_delta}
\end{equation}

with the additional bead factor $\delta_c^{+}$ from \eqref{Eq:beadfactor_c} where again the + in superscript indicates the assumption of a single plane wave for which $\delta_c^+$ is independent of position and direction. It is computed analytically from Mie theory, see equation \eqref{Eq:beadfactor_c_mie}. By fitting equation \eqref{Eq:fit_fun_delta} to the measured $\Delta \Gamma$, the true $E_{1D}$ can be extracted. A transfer matrix formalism \cite{madmax_theo_found} is used to relate the forward and backward amplitudes $E_{+}$ and $E_{-}$ with each other. In this case, by fixing the conductivity of the mirror to $\sigma_{Al}$, only the mirror position $z_m$ and forward amplitude $E_{+}$ are left as free parameters for the fit. Fitting is performed separately for each frequency using nonlinear least-squares. This means that the fit does not need to include complicated frequency dependencies as these will be absorbed into the fitted $E_{+}$. The fitted mirror position averaged over frequency is found to be $z_m = \SI{786.63(1)}{\milli\metre}$ showing both the accuracy of the bead pull method as well as the validity of plane wave propagation since deviations from plane wave propagation would likely manifest themselves as a fictitious frequency-dependence of physical lengths. The uncertainty on $E_{+}$ is given by the covariance matrix of the fit and is less than $\SI{8}{\percent}$.

The final step in calculating $P_{\mathrm{sig}}$ is to compute
\begin{equation}
    P_{\mathrm{sig}} = \frac{g_{a\gamma}^2\omega^2}{16 P_{\mathrm{in}}} \abs{\int_{-\infty}^{z_m} \dd{z} E_{1D} a B_e}^2.  
    \label{Eq:power_ax_1d}
\end{equation}
The external magnetic field and axion field $a B_e$ are set to one over the extent of the booster and decrease as $\sim \cos^2(\frac{z}{6\lambda})$ to zero outside of the booster. This makes sure that contributions to $P_{\mathrm{sig}}$ from magnetic field inhomogeneities remain negligible. Equation \eqref{Eq:power_ax_1d} is evaluated numerically. The resulting axion signal power of the dish antenna is shown in figure \ref{fig:mirror_bf}, parameterized as the boost factor $\beta^2$ to remove the dependency on $B_e$ and axion parameters. The uncertainty in $\beta^2$ is shown as a shaded band. The uncertainties of the input parameters to $E_{1D}$, namely $E_{+}$, $z_m$ and $\sigma_{Al}$, are propagated to $P_{\mathrm{sig}}$ using a Monte Carlo method. The uncertainty in $\alpha_0$ and $\delta_e$ directly translate to $P_{\mathrm{sig}}$ bringing the final uncertainty to less than $\SI{15}{\percent}$. Shown as well is the boost factor obtained from numerical simulation that does not include the antenna-booster resonances as well as a measurement where the antenna-booster resonances are filtered out using time gating. The baseline of measurement and numerical simulation both match, indicating a good calibration of the bead-pull method. Note that the reduction of the boost factor from an ideal value of one to $\sim 0.3$ is purely from the poor power coupling of the given Gaussian beam to the uniform $a {B}_e$ and not from losses. By increasing the waist size, one could obtain a theoretical maximum power coupling of $\sim 0.8$. Such a beam, however, would have a significant spill-over at the mirror contributing further to antenna-booster resonances. Even with the more conservative waist size of this setup, the antenna-booster resonances cause a maximum reduction of around $\SI{50}{\percent}$ from the baseline. While this is still a $\mathcal{O}(1)$ effect that is hardly visible in a logarithmic exclusion plot, one has to keep in mind that dish antennas do not have a lot of room for error and with a bit less care one could, at specific frequencies, end up with zero signal power. The upshot is that conversely, a decent boost in signal power is also possible, in this case of up to an order of magnitude. Overall, it should be noted that a realistic dish antenna has a far more interesting frequency response than it is usually assumed to have. 

\subsection{Dielectric haloscope}
\begin{figure}
    \centering
    \includegraphics[width=\textwidth]{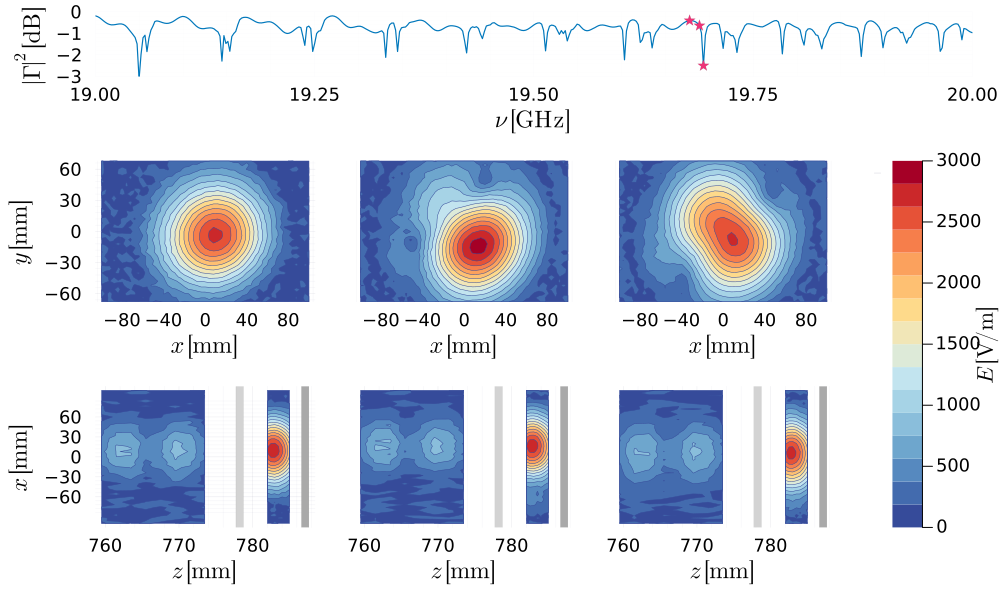}
    \caption{Electric field inside and in front of the dielectric haloscope. \textbf{Top panel}: Reflection coefficient of the setup. The markers indicate the frequencies at which the electric field is shown below from left to right. \textbf{Middle row}: Transverse field profile at $z\sim\SI{784}{\milli \metre}$, between disk and mirror. \textbf{Bottom row}: Longitudinal field profile between disk and mirror and in front of the disk at $y=\SI{0}{\milli\metre}$. The field is strongest between the disk (light gray band) and the mirror (dark gray band).}
    \label{fig:1_disk_E_evol}
\end{figure}

\FloatBarrier
To complete the minimal dielectric haloscope a sapphire disk is added in front of the mirror separated by mechanical spacers with a distance between disk and mirror of $d_v = \SI{7.95(5)}{\milli \meter}$. The disk has radius of $\SI{300}{\milli\metre}$ and a thickness of $d_{\epsilon}=\SI{1.00(5)}{\milli\metre}$. The relative permittivity of sapphire in the radial direction was measured to be $\epsilon = \num{9.3(1)}$ at \SI{20}{\giga\hertz} \cite{haotian2022}. The field is measured between the disk and mirror as well as in front of the disk. Figure \ref{fig:1_disk_E_evol} shows the field distribution of $|E|$ in the booster. The chosen frequencies are all on the booster resonance but not all on an antenna-booster resonance. In the left panel, on booster resonance but off antenna-booster resonance, the field is undisturbed. The middle and right panels show how the field is increasingly perturbed as the booster resonance and HoM antenna-booster resonance mix.

\begin{figure}
    \centering
    \begin{subfigure}{0.49\textwidth}
        \centering
        \includegraphics[width=\textwidth]{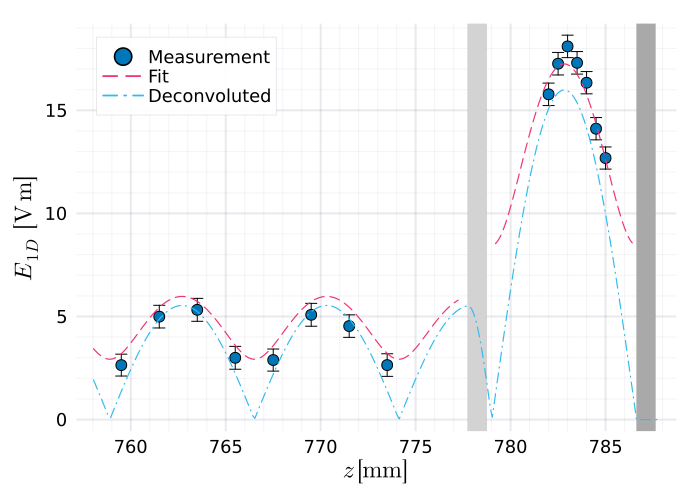}
        \caption{}
        \label{fig:1_disk_intdA_E}    
    \end{subfigure}
    \begin{subfigure}{0.49\textwidth}
        \centering
        \includegraphics[width=\textwidth]{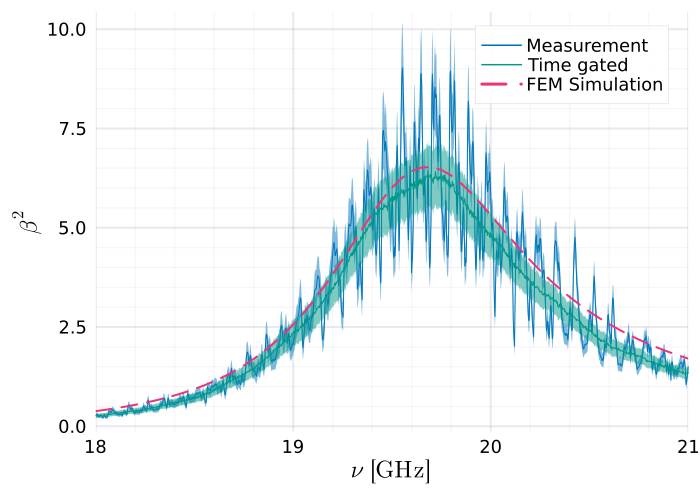}
        \caption{}
        \label{fig:1_disk_bf}    
    \end{subfigure}
    \caption{\textbf{(a)}: Transversely integrated electric field over $z$ at $\nu \sim \SI{19.7}{\giga\hertz}$. The measurements are fitted including finite size effects of the bead (dashed line) and then deconvolved (dashed-dotted line). \textbf{(b)}: Boost factor $\beta^2$ of the dielectric haloscope from measurement and simulation.}
    \label{fig:1_disk_integration}
\end{figure}

The procedure to calculate $P_{\mathrm{sig}}$ for the dielectric haloscope is analogous to the dish antenna case. As before, the measured $E$ is integrated in the transverse direction first to obtain $E_{1D}$ which is again fitted by equation \eqref{Eq:fit_fun_delta}. In the case of the minimal dielectric haloscope, the transfer matrix formalism relates the forward and backward amplitudes $E_{+}$ and $E_{-}$ not only in front of the plane mirror but also inside the disk and in front of the disk. For this, the additional fixed parameters are the permittivity and thickness of the disk. The plane mirror position is also fixed to the value obtained previously from the dish antenna setup. This leaves the position of the dielectric disk $z_\epsilon$ as well as the forward amplitude $E_{+}$ between disk and mirror as fit parameters. The measurement and fit are shown in figure \ref{fig:1_disk_intdA_E}. The fitted disk position varied less than $\SI{10}{\micro\metre}$ over frequency, again showing remarkable accuracy. As a cross-check, the plane mirror position is also fitted in which case it agrees with the dish antenna value within $\sim \SI{10}{\micro\metre}$. Thus plane wave propagation seems to be valid also for the minimal dielectric haloscope. As before, all the complicated effects over frequency and transverse directions are absorbed into $E_{+}$. The uncertainty of $E_{+}$ from the fit is less than $\SI{5}{\percent}$.

With $E_{1D}$ extracted from the fit, $P_{\mathrm{sig}}$ is calculated via equation \eqref{Eq:power_ax_1d} and shown as the boost factor in figure \ref{fig:1_disk_bf}. The uncertainty on $\beta^2$ is less than $\SI{22}{\percent}$, a bit higher than in the dish antenna case. This is attributed to the fact that more parameters enter the transfer matrix formalism, each with some uncertainty that is propagated to $\beta^2$ via Monte Carlo method. Shown as well is the result from the 2D axisymmetric FEM simulation and a time-gated measurement that, again, do not include the antenna-booster resonances which lead to a deviation of up to $\sim \SI{50}{\percent}$ between measurement and simulation. Overall, the effect of the antenna-booster resonances on $P_{\mathrm{sig}}$ seems tolerable. 
Extending the dielectric haloscope to more disks comes with its own challenge of increasingly tighter tolerances on geometric inaccuracies \cite{Knirck_2021}. In addition, it might not be practical to measure the electric field everywhere in a  many disks setup especially when considering a cryogenic setup. One might imagine a hybrid approach where a limited amount of bead pull measurements, for example in front of the booster, serve as input to a more sophisticated model capable of simulating the booster. This would still massively reduce computational complexity as the antenna and focusing mirror need not be included in the model while keeping complexity in the measurement manageable.

\FloatBarrier
\section{Conclusion}
\label{sec:conclusion}
This study applied a new approach to obtain the first realistic axion signal power of a dish antenna and dielectric haloscope. With the reciprocity approach, measurements of the reflection-induced electric field can be used to directly calculate the expected axion signal power with minimal model input. The electric field was measured with the bead pull method on three different setups. The first and simplest setup served as a benchmark test of the bead pull system and provided important input on the antenna, focusing mirror, and bead properties used in the other two setups. In the dish antenna and minimal dielectric haloscope, the baseline of the signal power agrees with the expectation from simulation within $\SI{10}{\percent}$. The antenna-booster resonances, however, add deviations of up to $\sim \SI{50}{\percent}$ from the baseline. The effect of the antenna-booster resonances can now, for the first time in any open haloscope, be fully accounted for. In particular, the reciprocity approach removes the need for a full simulation of the setup which is often not possible when considering optically large setups like dish antennas and dielectric haloscopes. Instead, only a simplified local model is needed to interpolate and deconvolve the measured electric field such that a more precise result can be obtained. In the future, this method will be applied on a larger dielectric haloscope capable of exploring new parameter space for axions.

\begin{acknowledgments}
The MADMAX collaboration made this work possible by providing the necessary hardware and valued feedback. Special thanks go to Olivier Rossel for his view from a unique vantage point with both useful and refreshing insights. This work is supported by the Deutsche Forschungsgemeinschaft (DFG, German Research Foundation) under Germany’s Excellence Strategy, EXC 2121, Quantum Universe (390833306). SK is supported by Fermi Research Alliance, LLC under Contract No. DE-AC02-07CH11359 with the U.S. Department of Energy, Office of Science, Office of High Energy Physics.
\end{acknowledgments}

\appendix
\FloatBarrier

\section{Signal processing}
\label{sec:appendix}
\begin{figure}
    \centering
    \begin{subfigure}{0.49\textwidth}
        \centering
        \includegraphics[width=\textwidth]{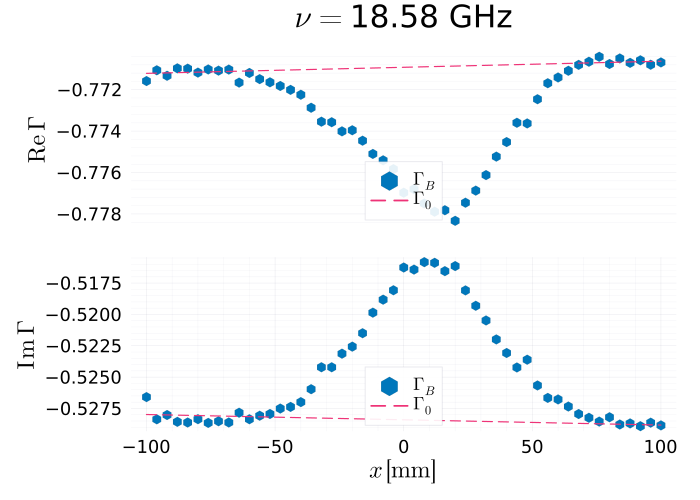}
        \caption{}
        \label{fig:sig_proc_gamma_fit}    
    \end{subfigure}
    \hfill
    \begin{subfigure}{0.49\textwidth}
        \centering
        \includegraphics[width=\textwidth]{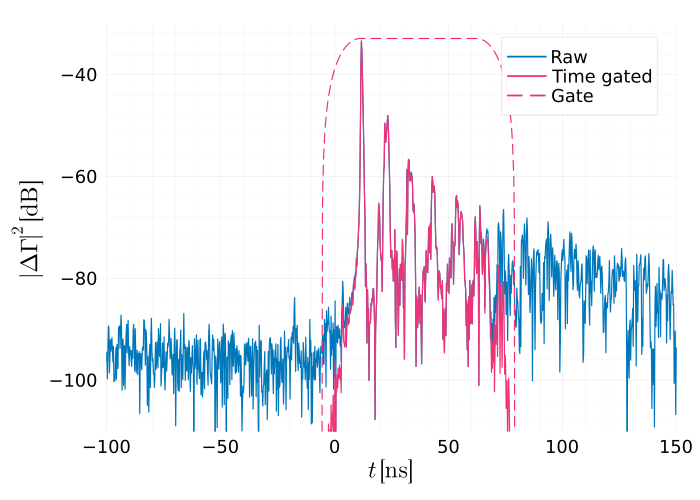}
        \caption{}
        \label{fig:sig_proc_delta_gamma_time}    
    \end{subfigure}
    \begin{subfigure}{0.49\textwidth}
        \centering
        \includegraphics[width=\textwidth]{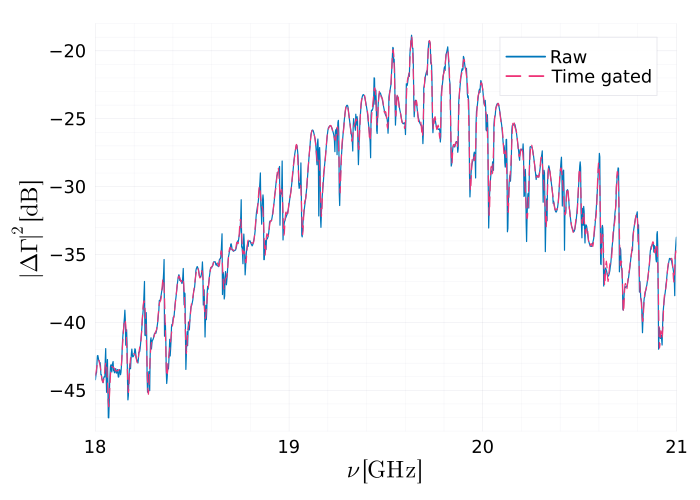}
        \caption{}
        \label{fig:sig_proc_delta_gamma_freq}    
    \end{subfigure}
    \hfill
    \begin{subfigure}{0.49\textwidth}
        \centering
        \includegraphics[width=\textwidth]{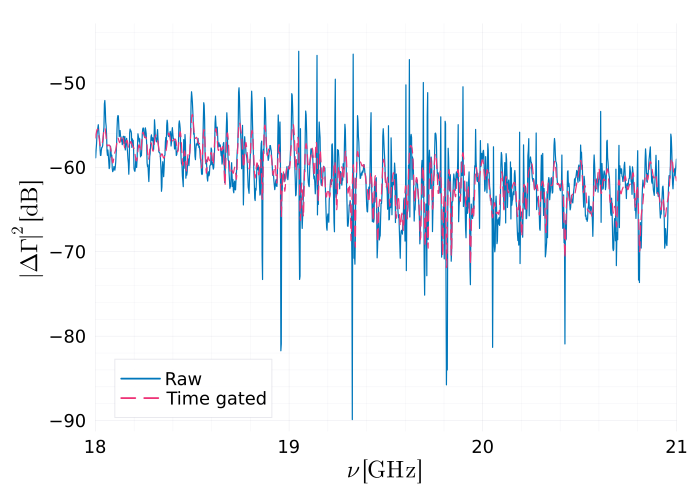}
        \caption{}
        \label{fig:sig_proc_delta_gamma_edge_freq}    
    \end{subfigure}
    \caption{Signal processing steps shown here on the dielectric haloscope setup. \textbf{(a)}: Real and imaginary part of $\Gamma$ over bead position $x$. The baseline reference without the bead $\Gamma_0$ is obtained from the measurements at the edges. \textbf{(b)}: Fourier transform of $\Delta\Gamma$. A gate is applied to filter out noisy components. \textbf{(c,d)}: Raw and time-gated $\Delta\Gamma$ over frequency at the beam's center (c) or outside the beam (d).}
    \label{fig:signal_processing}
\end{figure}

\begin{figure}
    \centering
    \includegraphics[width=\textwidth]{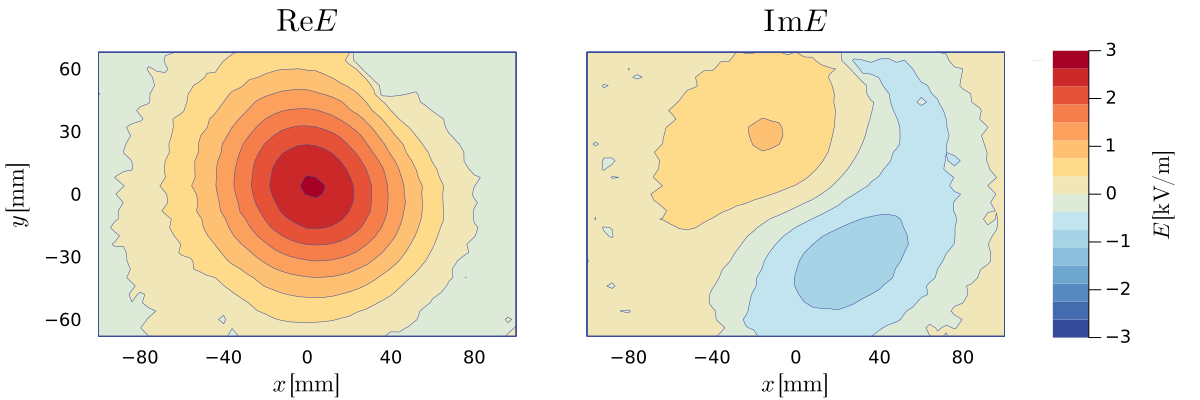}
    \caption{Real and imaginary part of the electric field between disk and mirror at \SI{19.88}{\giga\hertz}. }
    \label{fig:1_disk_E_complex_xyz_evol}
\end{figure}

The VNA measures the complex reflection coefficient $\Gamma_B(\nu,x,y,z)$ over frequency $\nu$ for each bead position $x,y,z$ resulting in a four-dimensional data set. This includes bead positions outside the beam that serve as the reference measurement without the bead $\Gamma_{0}$. Ideally, one would only need one reference measurement. However, since the setup is subject to long-term drifts due to changes in temperature, $\Gamma_0$ should be as close as possible in time to each $\Gamma_B$. This is achieved by always scanning across the full beam in the x-direction for each $y$ and $z$ position. Thus one obtains several reference measurements, left and right of the beam, for each $y$ and $z$ position. To account for drift during a scan in the x-direction, a linear fit between the left and right reference measurements interpolates $\Gamma_0$ for all $x$ positions. Figure \ref{fig:sig_proc_gamma_fit} shows $\Gamma_B$ and $\Gamma_0$ over $x$. Next, the difference is obtained,
\begin{equation}
    \Delta \Gamma = \Gamma_B - \Gamma_0 .
\end{equation}
Still, $\Delta \Gamma$ contains the effects of drift. This can be seen by Fourier transforming $\Delta \Gamma$ into the time domain shown in figure \ref{fig:sig_proc_delta_gamma_time}. The main peak corresponds to the bead position. Any feature earlier than that would ideally subtract out as it is unaffected by the bead. Removing these earlier features by time gating thus further reduces the effect of drifts and noise without losing information on ${E}$. Features after the main peak contain, however, the effect of additional antenna reflections and should ideally be preserved. Since each peak becomes successively smaller and noisier, the maximum number of peaks included in the time gate is six. This helps to further reduce noise without losing too much information on the antenna-booster resonances. To remove the antenna-booster resonance entirely, the gate would only include the main peak. Figure \ref{fig:sig_proc_delta_gamma_freq} shows $\Delta\Gamma$ in the beam's center before and after time gating. The difference is relatively small. Outside the beam, however, time gating helps to reduce noise significantly, as can be seen in figure \ref{fig:sig_proc_delta_gamma_edge_freq}. This is important because when calculating $E$ from $\Delta \Gamma$ via equation \eqref{Eq:nonresonant_small_steele}, one needs to take the square root of the complex $\Delta \Gamma$ and one can introduce phase jumps of $\pi$, which represents an unwrapping problem. The phase of $E$ is expected to vary slowly compared to the sampling density in the transverse $xy$ plane and a 2D unwrapping algorithm based on \cite{herraez2002} is applied separately to each $xy$ slice. Without any time gating, the algorithm often fails as noise makes unwrapping, particularly in two dimensions, notoriously difficult. As the axion signal power depends on the complex $E$ including phase, a failed unwrap can have a significant effect on $P_{\mathrm{sig}}$. Figure \ref{fig:1_disk_E_complex_xyz_evol} shows the unwrapped real and imaginary part of $E$ at a frequency with significant HoM contribution. 

\section{Bead factor from Mie theory}
\label{sec:app_mie}
Mie theory analytically treats the scattering of a plane wave off spherical objects. First, the exciting electric field $\vb*{E}$ and internal electric field $\vb*{E}_B$ are expanded in terms of vector spherical harmonics following standard procedure \cite{Bohren1998}. Using the orthogonality of vector spherical harmonics, the bead factors $\delta_e^+$ and $\delta_c^+$ can be expressed as 

\begin{equation}
    \delta_e^+ = \sum_{n=1}^{\infty} (-1)^n \left(K_n c_n - L_n d_n\right),
    \label{Eq:beadfactor_mie}
\end{equation}
\begin{equation}
    \delta_c^+ = \sum_{n=1}^{\infty}  \left(K_n c_n + L_n d_n\right).
    \label{Eq:beadfactor_c_mie}
\end{equation}
The coefficients $c_n$, $d_n$ are
\begin{equation}
c_n = \frac{j_n(x)[xh_n^{(1)}(x)]' - h_n^{(1)}(x)[xj_n(x)]'}
{j_n(mx)[xh_n^{(1)}(x)]' - h_n^{(1)}(x)[mxj_n(mx)]'}      
\end{equation}
and
\begin{equation}
d_n = \frac{mj_n(x)[xh_n^{(1)}(x)]' - mh_n^{(1)}(x)[xj_n(x)]'}
{m^2 j_n(mx)[xh_n^{(1)}(x)]' - h_n^{(1)}(x)[mxj_n(mx)]'},      
\end{equation}
with the spherical Bessel of the first kind $j_n$, first spherical Hankel function $h_n^{(1)}$, size parameter $x=k r_B$, and relative refractive index $m=\sqrt{\epsilon_B}$. The prime denotes differentiation with respect to the argument in parentheses. The coefficients $K_n$ and $L_n$ are

\begin{equation}
    K_n = \frac{2n+1}{m^2-1}x^2 \left[j_{n-1}(x)j_n(mx) - m j_n(x)j_{n-1}(mx)\right]
\end{equation}
and
\begin{equation}
\begin{aligned}
     L_n &= \frac{x^2}{m^2-1} \{(n+1)[j_{n-2}(x)j_{n-1}(mx) - m j_{n-1}(x)j_{n-2}(mx)]\\&+ n[j_n(x)j_{n+1}(mx) - m j_{n+1}(x)j_n(mx)]\}.   
\end{aligned}
\end{equation}

The bead factors are then computed numerically in Julia using the SpecialFunctions.jl package\footnote{https://specialfunctions.juliamath.org/stable/} and have been checked against COMSOL FEM simulations.


\FloatBarrier


\bibliographystyle{JHEP}
\bibliography{bibliography.bib}

\end{document}